\documentclass[prb, twocolumn, superscriptaddress, showpacs,floatfix]{revtex4-1}
\usepackage{hyperref}
\usepackage{amssymb}
\usepackage{amsmath}
\usepackage{float}
\usepackage{bbm}
\usepackage{graphicx}
\usepackage{epsfig}
\usepackage{epstopdf}

\begin{document}

\title{Noise-compensating pulses for electrostatically controlled silicon spin qubits}
\author{Xin Wang}
\affiliation{Condensed Matter Theory Center, Department of Physics, University of Maryland, College Park, MD 20742, USA}
\author{F.~A.~Calderon-Vargas}
\affiliation{Department of Physics, University of Maryland Baltimore County, Baltimore, MD 21250, USA}
\author{Muhed S. Rana}
\affiliation{Department of Physics, University of Maryland Baltimore County, Baltimore, MD 21250, USA}
\author{J.~P.~Kestner}
\affiliation{Department of Physics, University of Maryland Baltimore County, Baltimore, MD 21250, USA}
\author{Edwin Barnes}
\affiliation{Condensed Matter Theory Center, Department of Physics, University of Maryland, College Park, MD 20742, USA}
\affiliation{Joint Quantum Institute, University of Maryland, College Park, MD 20742, USA}
\author{S.~Das Sarma}
\affiliation{Condensed Matter Theory Center, Department of Physics, University of Maryland, College Park, MD 20742, USA}
\affiliation{Joint Quantum Institute, University of Maryland, College Park, MD 20742, USA}
\date{\today}

\begin{abstract}
We study the performance of {\sc supcode}---a family of dynamically correcting pulses designed to cancel simultaneously both
Overhauser and charge noise for singlet-triplet spin qubits---adapted to silicon devices with electrostatic control. We consider both natural Si and isotope-enriched Si systems, and in each case we investigate the behavior of individual gates under static noise and perform randomized benchmarking to obtain the average gate error under realistic $1/f$ noise. We find that in most cases {\sc supcode} pulses offer roughly an order of magnitude reduction in gate error, and especially in the case of isotope-enriched Si, {\sc supcode} yields gate operations of very high fidelity. We also develop a version of {\sc supcode} that cancels the charge noise only,  ``$\delta J$-{\sc supcode}'', which is particularly beneficial for isotope-enriched Si devices where charge noise dominates Overhauser noise, offering
a level of error reduction comparable to the original {\sc supcode} while yielding gate times that are $30\%$ to $50\%$ shorter. Our results show that the {\sc supcode} noise-compensating pulses provide
a fast, simple, and effective approach to error suppression, bringing gate errors well below the quantum error correction threshold in principle.

\end{abstract}



\maketitle

Ever since Kane's original proposal for a silicon-based quantum computer,\cite{Kane.98} Si has been a prime candidate as a host material for solid-state quantum computing.\cite{Zwanenburg.13} A decisive advantage of Si is that it is available in an isotope-enriched form ($^{28}$Si) that has zero nuclear spin. This allows for the nearly complete removal of decoherence due to the hyperfine interaction between the qubit and surrounding nuclear spins (Overhauser noise), leading to remarkably long coherence times and high control fidelity.\cite{Morton.08,Morley.10,Tyryshkin.11,Pla.12,Pla.13,Wolfowicz.13,Muhonen.14,Kim.14,Wolfowicz.14} Recent years have witnessed substantial experimental progress in the fabrication, initialization, readout and control of spins in both phosphorous donors and gate-defined quantum dots in Si
systems.\cite{Slinker.05,Lim.09,Morello.10,Shi.11,Simmons.11,Prance.12,Maune.12,Shi.13,Dehollain.14,Wu.14,Muhonen.14,Kim.14,Wolfowicz.14} For a P donor qubit on Si, both the donor electron spin\cite{Pla.12} and the nuclear spin\cite{Pla.13} have been explored as potential qubits, exhibiting coherence times up to 30 seconds and control fidelities exceeding 99.99\% in recent experiments operating the qubit as a quantum memory,\cite{Muhonen.14} both of which are the highest achieved for any solid state qubit. In laterally defined Si quantum dot systems, coherent operation of a singlet-triplet spin qubit\cite{Levy.02,Petta.05} has also been demonstrated.\cite{Maune.12,Wu.14}

While Si qubits have been shown repeatedly to possess long information storage (i.e. quantum memory) times, it remains an open question on how to extend this long-lived coherence to qubits undergoing quantum gate operations. Experiments have progressed to such a stage where high fidelity quantum gates are now within reach. While quantum devices built on isotope-enriched Si already enjoy many advantages, some noise remains. In particular, the residual $^{29}$Si impurities
distort the wave function of the donor qubit or
the lateral confinement potential of the quantum dot, which either broadens the ESR resonance in the former case\cite{Feher.59,Morley.10} or shifts the exchange interaction in the latter,\cite{Wu.14} both causing inaccuracies in control. More importantly,
additional errors can arise through electrical fluctuations from either the leads or nearby charged impurities, and from variations in control parameters.\cite{Li.10,Rahman.12,Wu.14} While a vast literature on dynamically corrected gates (DCGs) has existed long before the conception of spin qubits, it was only recently realized that the special constraints imposed by solid state spin systems often require brand-new approaches to developing DCGs.\cite{Wang.12,Khodjasteh.12,Kosut.13,Cerfontaine.14,Kestner.13,Wang.14}
While one of these approaches, known as {\sc supcode},\cite{Wang.12} has already been experimentally demonstrated for nitrogen-vacancy centers in diamond\cite{Rong.14} and is also applicable to single donor spins in Si, it is particularly suitable for singlet-triplet qubits subject to Overhauser and charge noise
because it is designed to yield comparatively simple pulse sequences that cancel both while respecting all experimental constraints, including the restriction to positive, single-axis control. In the case of isotope-enriched Si, however, canceling Overhauser noise is no longer necessary, leaving open the possibility that robust quantum operations on Si devices might be realized with even simpler and faster control sequences.

In this paper, we show that {\sc supcode} pulses can substantially improve gate fidelities in silicon singlet-triplet spin qubits.
We begin by quantifying the extent to which charge noise corrupts the application of individual quantum gates and demonstrate the cancelation of these errors effected by {\sc supcode} pulses. We then provide a more systematic analysis of the performance of {\sc supcode} by studying its effectiveness for long sequences of quantum gates subject to $1/f$ noise, which describes both Overhauser and charge noise.\cite{Dial.13} Through randomized benchmarking,\cite{magesan_characterizing_2012} we extract the average error per gate for the set of single-qubit Clifford gates, and we find that {\sc supcode} pulses remain robust against noise in this realistic situation. We also develop a version of {\sc supcode} that works particularly well for isotope-enriched Si while being about 30\% to 50\% faster, and we demonstrate that it offers a level of error reduction comparable to the original version. Additionally, we discuss the applicability of {\sc supcode} to $1/f$ noise with different exponents. These results illustrate that Si quantum devices can achieve long coherence times not only when the qubit is idle but even while it is undergoing operations.

We start with the model Hamiltonian governing a singlet-triplet qubit,\cite{Levy.02,Petta.05,Wang.12} which can be realized in Si either by gating exchange-coupled P donors\cite{Kalra.14,Weber.14} or by a gate-defined double dot in a heterostructure:\cite{Maune.12,Wu.14}
\begin{equation}\label{eq:ham}
H(t)=\frac{h}{2}\sigma_x+\frac{J\left[\epsilon\left(t\right)\right]}{2}\sigma_z.
\end{equation}
The computational bases are
$|0\rangle=|\mathrm{T}\rangle=({|\!\uparrow\downarrow\rangle}+{|\!\downarrow\uparrow\rangle})/\sqrt{2}$ and $|1\rangle=|\mathrm{S}\rangle=\left({|\!\uparrow\downarrow\rangle}-{|\!\downarrow\uparrow\rangle}\right)/\sqrt{2}$, where ${|\!\downarrow\uparrow\rangle}=c_{1\downarrow}^\dagger c_{2\uparrow}^\dagger|\mathrm{vacuum}\rangle$ with $c_{j\sigma}^\dagger$ creating an electron with spin $\sigma$ in the $j$th dot.  Rotations around the $x$-axis are performed with a magnetic field gradient across the double-dot system, which in energy units reads $h=g\mu_B\Delta B_z$. This magnetic field gradient can be generated by a micromagnet, as demonstrated in Si/SiGe quantum dot systems,\cite{Wu.14} where $h$ is approximately $30-60$ neV.
Rotations around the $z$-axis are done using the exchange interaction $J$, the energy level splitting between $|\mathrm{S}\rangle$ and $|\mathrm{T}\rangle$. The magnitude of $J$ is controlled by the detuning $\epsilon$, namely the tilt of the double-well confinement potential, through the gate voltages. Experimentally, $J$ can be tuned from close to zero up to tens of $\mu$eV.\cite{Shi.11,Maune.12,Kalra.14,Wu.14} One therefore has fast, all-electrical control over the rotation rate around the $z$-axis. On the other hand, the magnetic field gradient $h$ cannot be changed efficiently during a gate operation, and we will take it to be constant throughout the computation.

In general, decoherence can arise through both $h$ and $J$. Fluctuations arising from Overhauser noise add a small, but unknown error term $\delta h$ to the Hamiltonian: $h\rightarrow h+\delta h$.  For natual Si, $\delta h$ is on the scale of neV, \cite{Wu.14, Kalra.14} corresponding to a few percent of $h$. However, for isotope-enriched silicon, $\delta h$ can be reduced by up to three orders of magnitude,\cite{Kalra.14}   down to $\delta h/h\sim10^{-5}$.
On the other hand, impurities on the substrate lead to deformations of the confinement potential and consequently the energy level structure, thereby causing charge noise.\cite{Li.10,Rahman.12,Wu.14} This noise produces fluctuations in the exchange energy, $\delta J$, through fluctuations in the detuning $\epsilon$. In this work, we assume a phenomenological form of $J=J_0\exp(\epsilon/\epsilon_0)$, implying $\delta J=J\delta\epsilon/\epsilon_0$,\cite{Shulman.12,Dial.13,Wu.14} but our method applies to other forms also.\cite{Wang.14}  Estimates of $\delta\epsilon/\epsilon_0$ vary, ranging from 10\% ($\delta\epsilon\sim\mu$eV) \cite{Wu.14} to $10^{-4}$ (cf. Ref.~\onlinecite{Culcer.09}).  We study a range of $\delta\epsilon$ values in this work.

We previously developed a family of composite pulses ({\sc supcode}) that cancels both Overhauser noise\cite{Wang.12} and charge noise.\cite{Kestner.13, Wang.14} The basic idea is as follows: assuming the noise is quasi-static, one expands the evolution operator to first order in both $\delta h$ and $\delta\epsilon$. In order to cancel these error terms, the original rotation is supplemented by an (imperfect) identity operation, carefully designed such that the error arising from performing this identity operation exactly cancels that of the target rotation. This formalism reduces to a multi-dimensional optimization problem involving six coupled nonlinear equations (corresponding to error terms for $\sigma_x$, $\sigma_y$ and $\sigma_z$, each of which has $\delta h$ and $\delta\epsilon$ components), from which the pulse sequence can be obtained.  A detailed description of the theoretical framework is presented in Ref.~\onlinecite{Wang.14}, and we shall refer to these sequences as ``full {\sc supcode}'' sequences in this work because they cancel both Overhauser noise and charge noise.

\begin{figure}[t]
    \centering
    \includegraphics[width=8cm, angle=0]{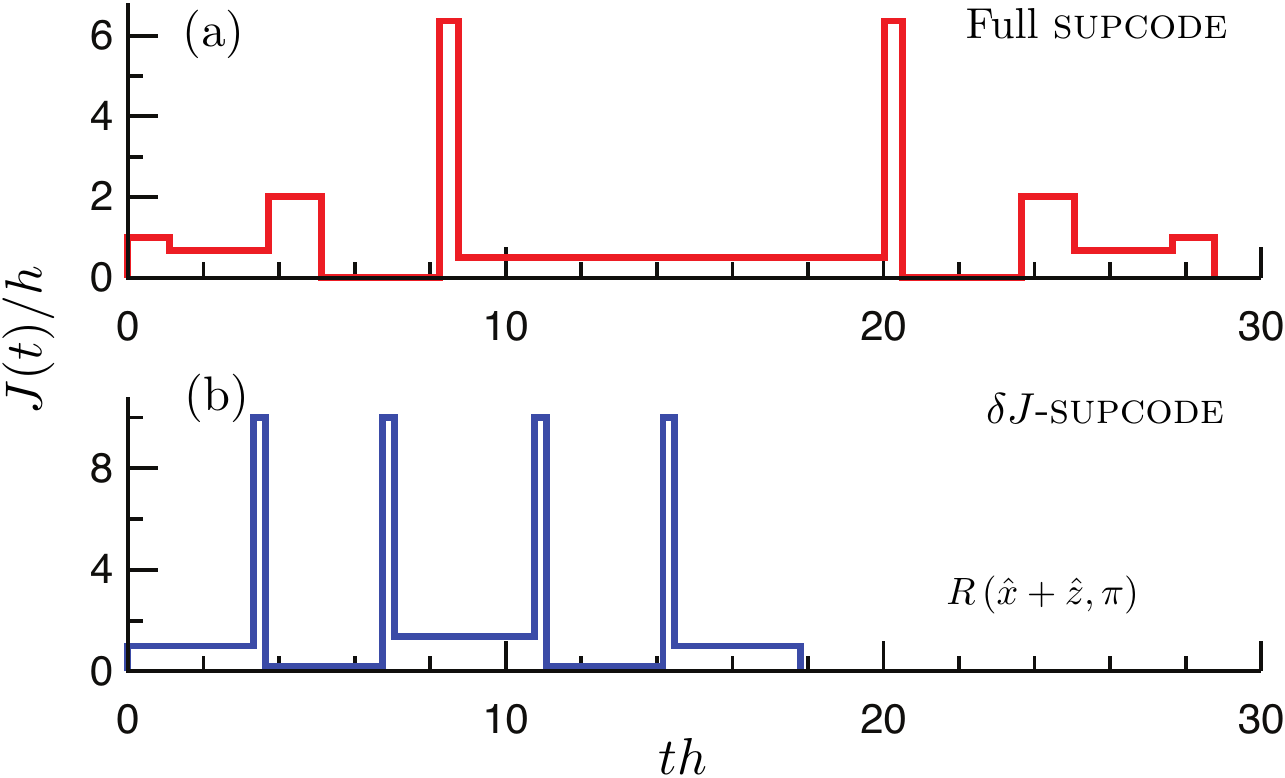}
    \caption{Pulse shape for the Hadamard gate, $R(\hat{x}+\hat{z},\pi)$. (a) Full {\sc supcode} canceling both Overhauser and charge noise. (b) $\delta J$-{\sc supcode} canceling charge noise only.}
    \label{fig1}
\end{figure}

While undoubtedly full {\sc supcode} works for the problem at hand, further optimizations are possible for an isotope-enriched Si system. Because the Overhauser noise is negligible in this case, one can completely ignore the $\delta h$ component of the error, leading to a simplification of the problem since only three coupled nonlinear equations need to be solved, and the resulting sequences are much shorter and simpler. In Fig.~\ref{fig1} we show pulse profiles comparing the full {\sc supcode} and $\delta J$-{\sc supcode}
for a Hadamard gate. Detailed sequences and pulse parameters for each of the single-qubit Clifford gates are given in the Appendix. In obtaining these results, we restricted the strength of the exchange to $J\le 10h$ so that the resulting sequences are more feasible for experimental implementation. In the Appendix we give explicit numerical results of the gate durations of $\delta J$-{\sc supcode}, and also show that $\delta J$-{\sc supcode} is about 30\% to 50\% shorter than the full {\sc supcode}.
We shall also show, in the remainder of this paper, that for parameter regimes of interest, the error reduction of $\delta J$-{\sc supcode} is comparable to that of full {\sc supcode} for isotope-enriched Si, making it particularly suitable for experimental realization in these systems.

\begin{figure}[t]
    \centering
    \includegraphics[width=8cm, angle=0]{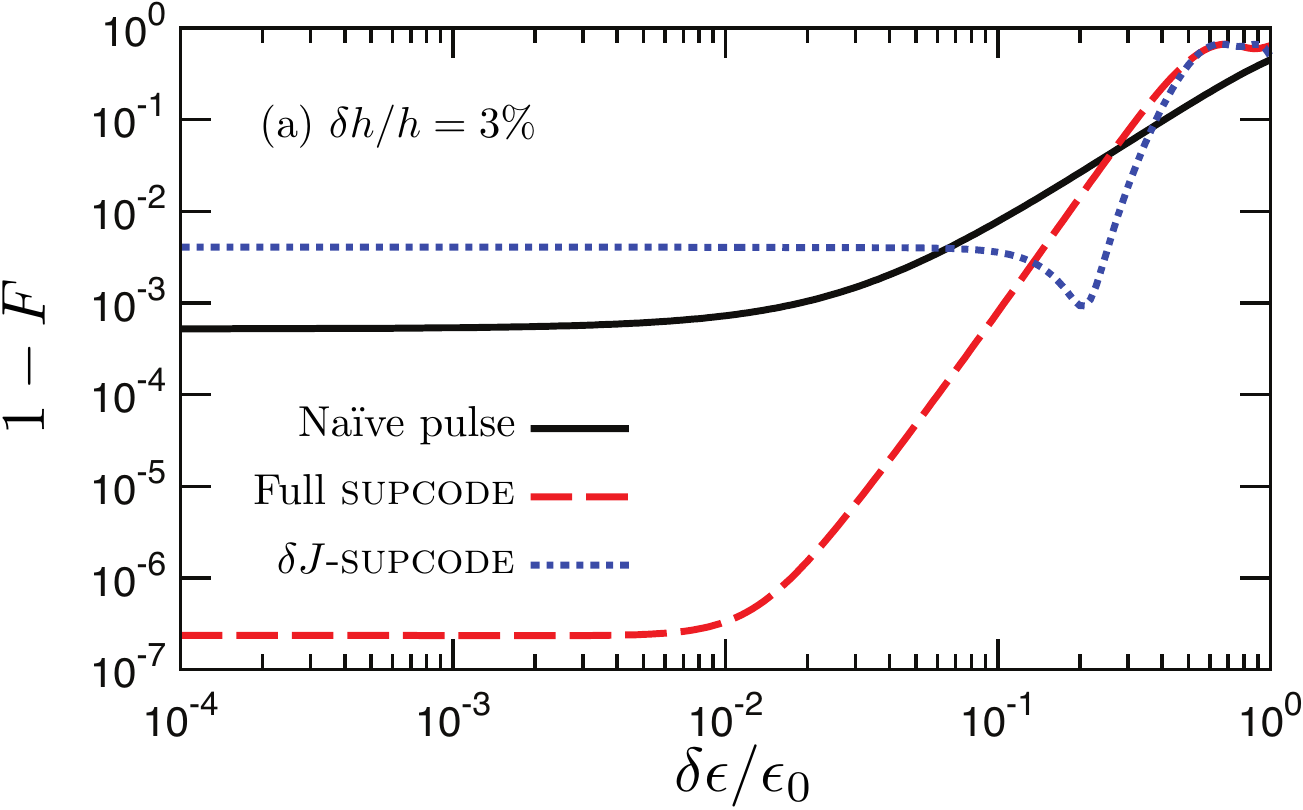}
    \includegraphics[width=8cm, angle=0]{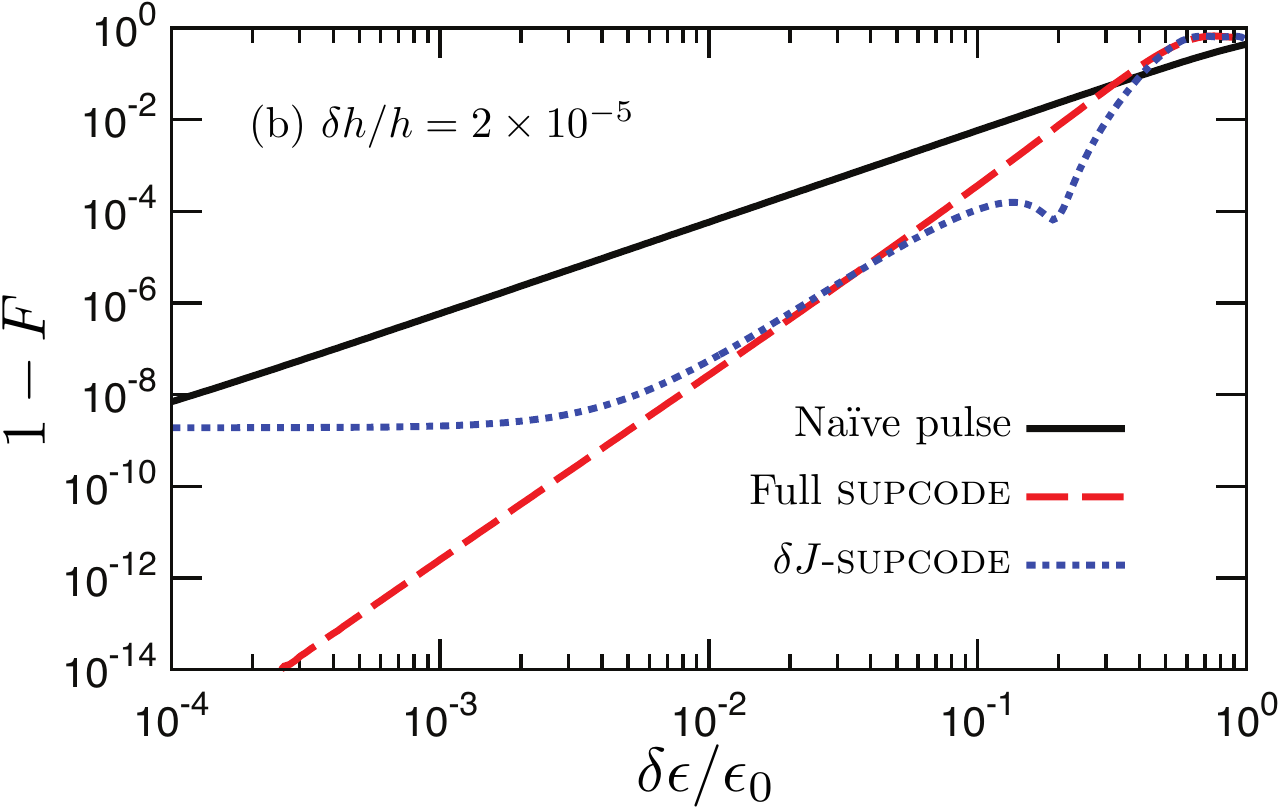}
    \caption{Infidelity of the Hadamard gate vs. error in detuning (charge noise) $\delta\epsilon/\epsilon_0$ for (a) $\delta h/h=3\%$ relevant to natual Si and (b) $\delta h/h=2\times10^{-5}$ relevent to isotope-enriched Si. Infidelity is shown for the na\"ive pulse (black solid), full {\sc supcode} (red dashed), and $\delta J$-{\sc supcode} (blue dotted). Note the different scales of the $y$-axis.}
    \label{fig2}
\end{figure}

Figure~\ref{fig2} shows the infidelity
(cf.~Ref.~\onlinecite{Bowdrey.02}) of the Hadamard gate versus quasi-static charge noise strength. We model natural Si as having $\delta h/h=3\%$, while isotope-enriched Si has $\delta h/h=2\times 10^{-5}$, consistent with experimental values.\cite{Wu.14,Kalra.14} Fig.~\ref{fig2}(a) shows the result for natural Si. While the curve for the na\"ive pulse saturates at an infidelity of about $10^{-3}$, the full {\sc supcode} has a residual error of only $10^{-7}$ for charge noise of less than $1\%$, and if the quantum error correction threshold is set at $10^{-4}$, the range of charge noise that can be tolerated is as large as 5\%. This is as expected because full {\sc supcode} cancels both $\delta h$ and $\delta\epsilon$ errors. Turning to the curve for $\delta J$-{\sc supcode}, we see that for most of the $\delta\epsilon$ range ($\delta\epsilon\lesssim5\%$), its performance is worse than the na\"ive pulse. This is understandable because $\delta J$-{\sc supcode} completely ignores the Overhauser noise, and this is precisely the range where Overhauser noise is dominant.

Figure~\ref{fig2}(b) shows the case of isotope-enriched Si. From the different scale of the $y$-axis compared with Fig.~\ref{fig2}(a) one immediately notices that the gate error is substantially reduced for all three curves as expected for negligible Overhauser noise. The exceedingly low errors achieved by the full {\sc supcode} are consistent with the remarkably long coherence times measured in experiments.\cite{Muhonen.14} At a typical quantum error correction threshold of $10^{-4}$, the na\"ive pulse tolerates charge noise up to $1\%$, whereas the full {\sc supcode} tolerates as much as $7\%$. While this range does not differ from the case of natural Si, the gate error for charge noise less than $10^{-2}$ is substantially smaller due to the absence of Overhauser noise.
Note that for the isotope-enriched Si, $\delta J$-{\sc supcode} has low error comparable with full {\sc supcode} for a wide range of noise amplitudes. Even in the regime where it is not as effective as full {\sc supcode} ($\delta\epsilon/\epsilon_0\lesssim1\%$), the error is still greatly reduced compared to that of the na\"ive pulse.
Similar behavior is observed for all single-qubit Clifford gates (not shown here) with the exception of $x$-rotations, where $\delta J$-{\sc supcode} has a constant error because charge noise does not come into play.

While individual gate errors such as those shown in Fig.~\ref{fig2} are strongly indicative of the improvements afforded by {\sc supcode}, a more complete characterization of this improvement is obtained by studying the error over long sequences of gates through randomized benchmarking.\cite{magesan_characterizing_2012}
We have so far assumed a static noise model, which is a good approximation for single gates,\cite{Wang.14} but a more realistic model taking into account temporal variations of the noise is needed for longer gate sequences. To include these effects, we employ a $1/f$ noise model in which both the Overhauser and charge noise have a power spectrum of the form $S_k(\omega)=\beta_k/\omega^\gamma$, where $k\in\{h, \epsilon\}$ denotes Overhauser $(h)$ and charge noise $(\epsilon)$, and $\beta_h$ and $\beta_\epsilon$ denote the noise amplitudes. We choose the exponent to be $\gamma=1.5$, which is consistent with the upper bound estimated from experiments.\cite{Pla.13}
We have considered other values of $\gamma$, and the results will be presented later in the paper.
We implement randomized benchmarking by averaging the fidelity over random sequences of single-qubit Clifford gates and over different noise realizations for a varying number of gates in the sequences. The fidelity of a sequence behaves as  $\frac{1}{2}\left[1+(1-2d)^n\right]$, where $d$ is the average error per gate, and $n$ is the number of Clifford gates applied.\cite{Knill.08} We fit our results to the exponentially decaying function $\frac{1}{2}\left(1+e^{-\gamma n}\right)$, where the average error per gate is related to the fitted exponent via $d=\left(1-e^{-\gamma}\right)/2$. The realizations of $1/f$ noise are generated using a weighted sum of random telegraph signals.\cite{kogan, Wang.14}

\begin{figure}[t]
    \centering
    \includegraphics[width=8cm, angle=0]{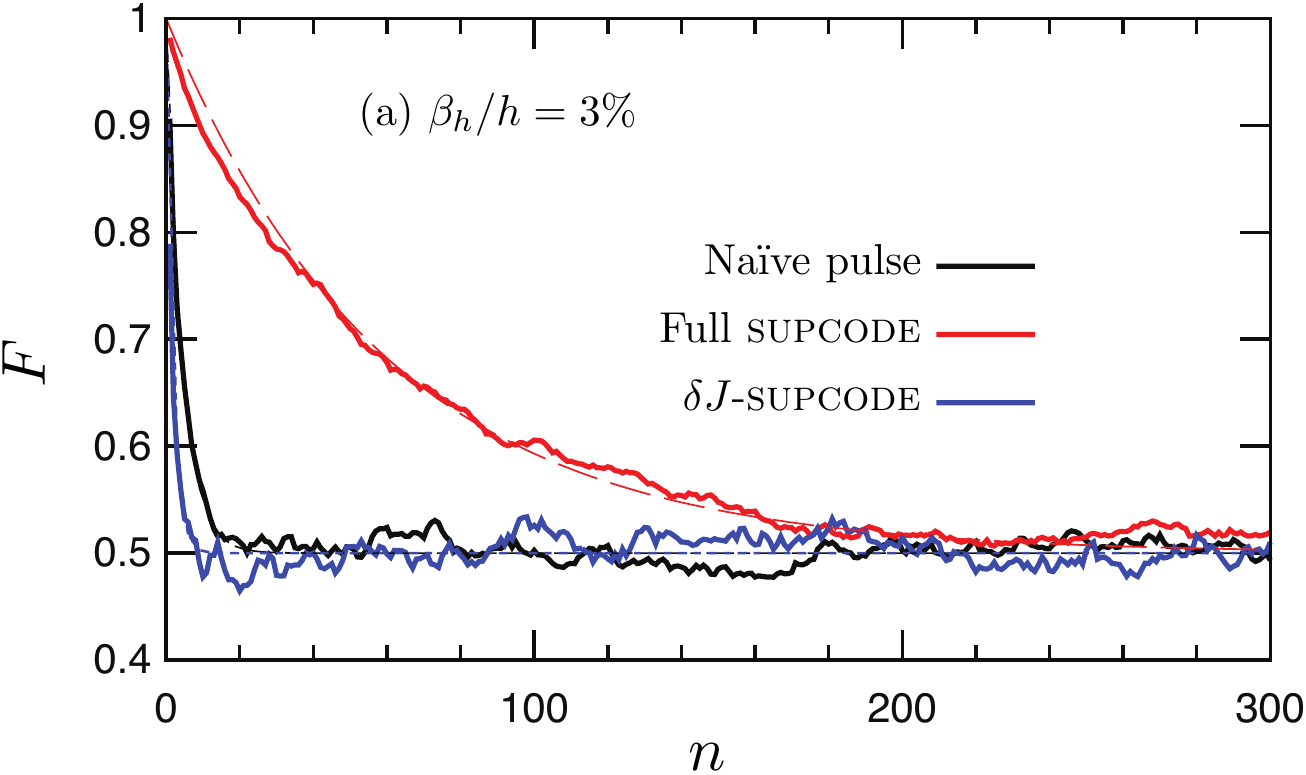}
    \includegraphics[width=8cm, angle=0]{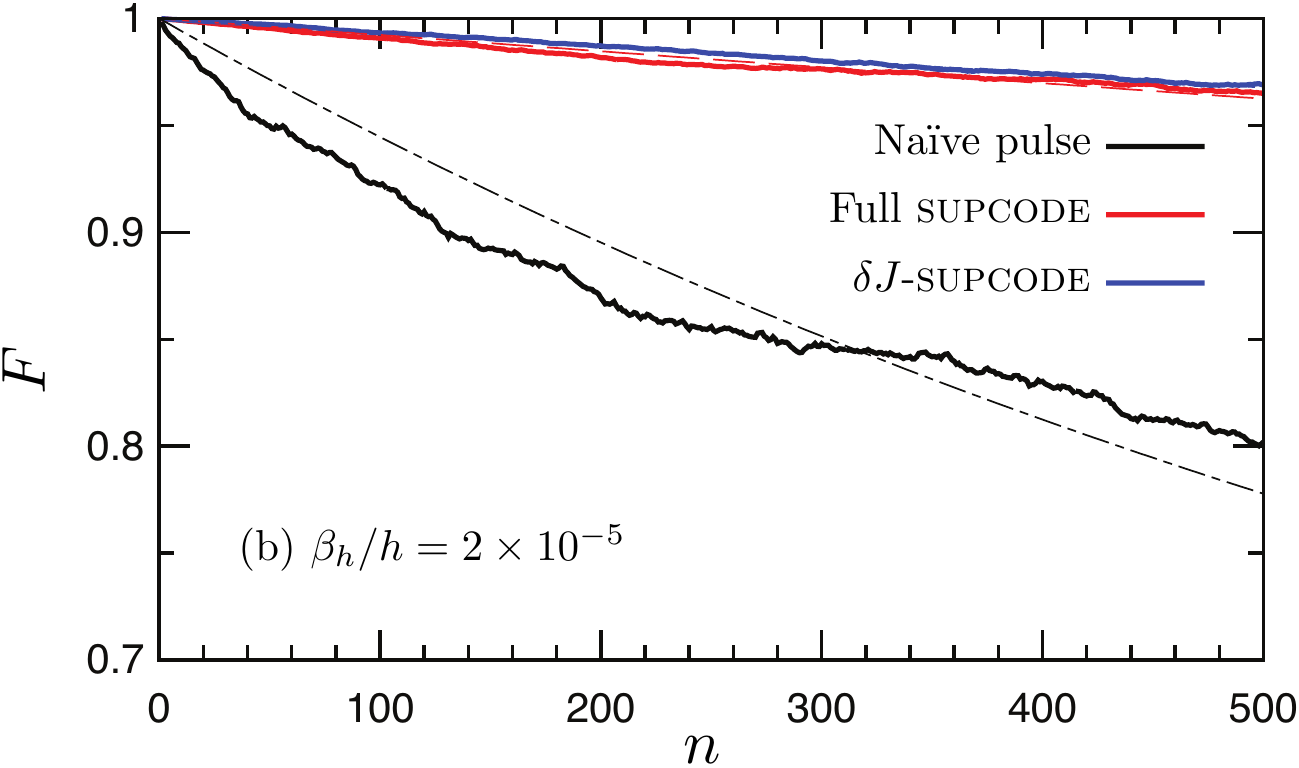}
    \caption{Decay of the gate fidelity via randomized benchmarking under $1/f$ noise for charge noise with amplitude $\beta_\epsilon/\epsilon_0=1\%$ for  (a) $\beta_h/h=3\%$ relevant to natual Si and (b) $\beta_h/h=2\times10^{-5}$ relevant to isotope-enriched Si. The na\"ive pulse, full {\sc supcode} and $\delta J$-{\sc supcode} are shown as black, red and blue lines respectively, and the thin black dot-dashed, red dashed and blue dotted lines are fits of the corresponding data to $\frac{1}{2}(1+e^{-\gamma n})$. Note the different scales of both the $x$ and $y$ axes.}
    \label{fig3}
\end{figure}

\begin{figure}[t]
    \centering
    \includegraphics[width=8cm, angle=0]{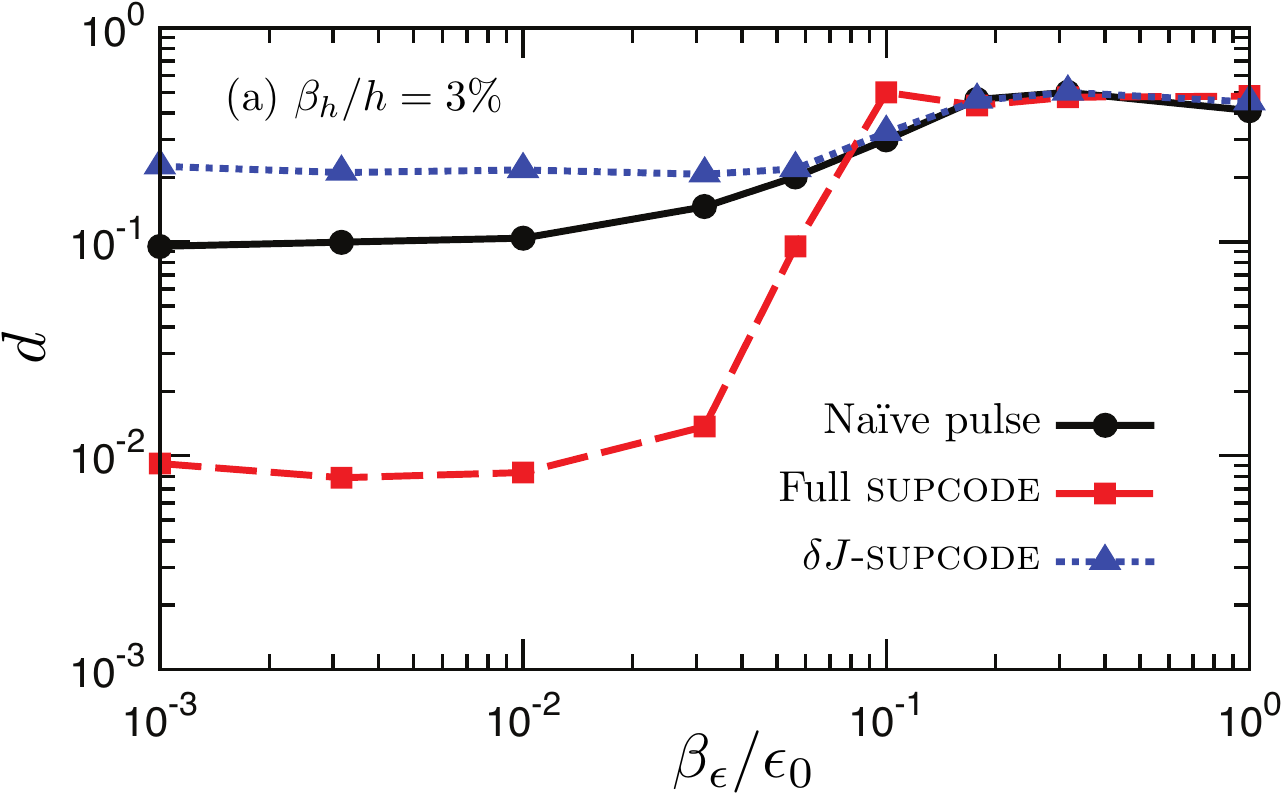}
    \includegraphics[width=8cm, angle=0]{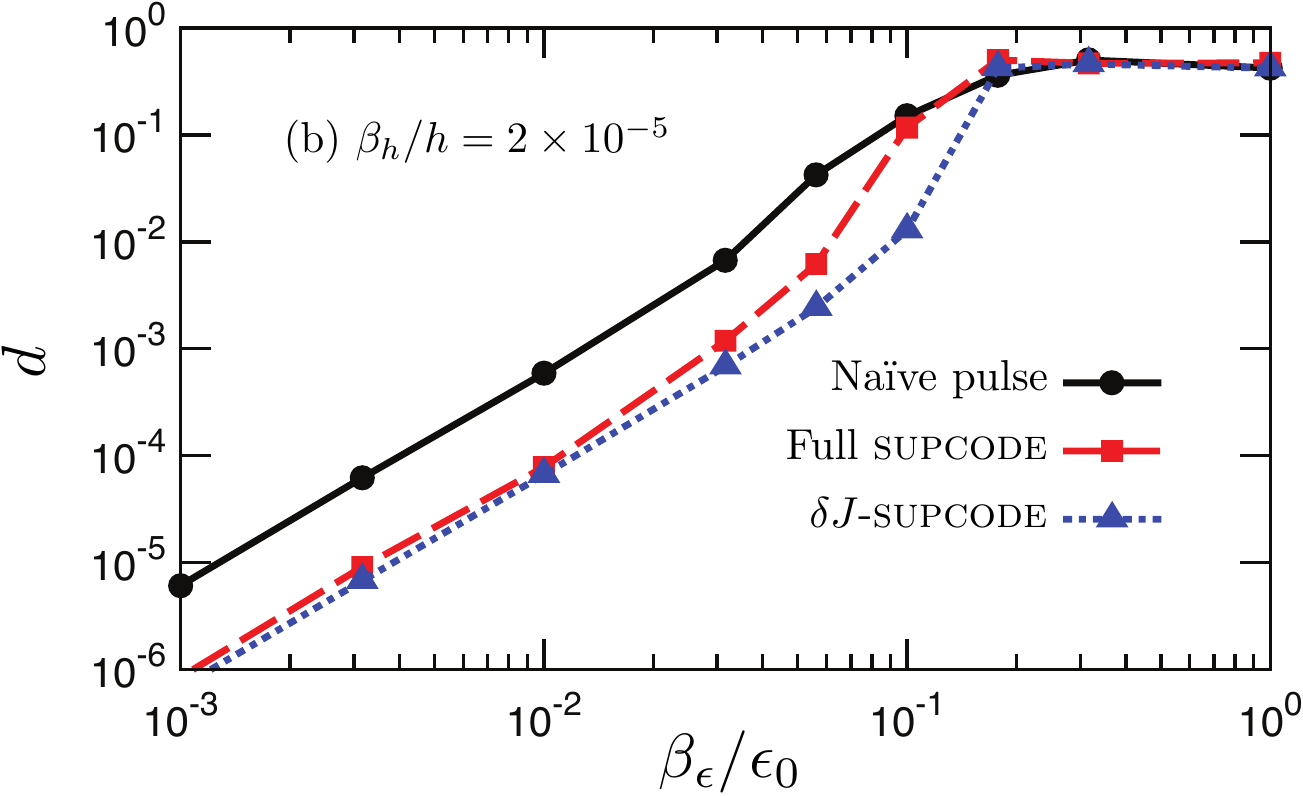}
    \caption{Average error per gate vs. amplitude of the charge noise $\beta_\epsilon/\epsilon_0$ for  (a) $\beta_h/h=3\%$ relevant to natual Si and (b) $\beta_h/h=2\times10^{-5}$ relevant to isotope-enriched Si. The gate error is shown for the na\"ive pulses (black solid), full {\sc supcode} (red dashed), and $\delta J$-{\sc supcode} (blue dotted). Note the different scales of the $y$-axis.}
    \label{fig4}
\end{figure}

Figure~\ref{fig3} shows the randomized benchmarking results for charge noise with $\beta_\epsilon/h=1\%$ (note that one should not directly compare this result to that of constant noise with $\delta\epsilon/\epsilon_0=1\%$ since the noise spectra are completely different). Fig.~\ref{fig3}(a) shows the result for $\beta_h/h=3\%$, relevant to natural Si. These results are generated using the average of 200 independent realizations of noise/gate sequences, each of which contains up to 300 Clifford gates. All curves asymptote to fidelity $1/2$ as expected, but the na\"ive pulse and $\delta J$-{\sc supcode} 
drops to $1/2$ much more steeply, reaching the asymptotic value after less than 20 gate operations.  Full-{\sc supcode} pulses, on the other hand, maintain the fidelity much longer, with 70\% fidelity after 60 gate operations. This clearly demonstrates that {\sc supcode} pulses play a crucial role in error reduction.  The result is consistent with that of Fig.~\ref{fig2} in that when nuclear noise is present, $\delta J$-{\sc supcode} does not work well, even worse than a na\"ive pulse. Turning to Fig.~\ref{fig3}(b), the results for $\beta_h/h=2\times10^{-5}$ relevant to isotope-enriched Si, we see a huge reduction in error. Here we average over 100 independent realizations of the noise and gate sequences and increase the number of gates in each sequence to 500.  We see that even with the na\"ive pulses, the fidelity remains larger than 80\% after 400 gate operations. Most remarkably, when {\sc supcode} pulses are used, the fidelity remains greater than 95\% even after 500 gate operations. This fact reinforces the main point of this paper: while isotope-enriched Si is already superior to natural Si, application of {\sc supcode} reduces the error much further.
We see again that $\delta J$-{\sc supcode} has an error reduction ability comparable with the full {\sc supcode}, so it is particularly useful for isotope-enriched Si since it is typically shorter, less complicated, and works almost perfectly when charge noise plays a much more significant role than Overhauser noise.

We have also performed randomized benchmarking for different magnitudes of the charge noise, and these results are summarized in Fig.~\ref{fig4}, where we show the average error per gate, $d$, obtained from the fitted decaying exponent. We note from Fig.~\ref{fig4}(a) that, for charge noise below 3\% in natural Si, the na\"ive pulse has a gate error of approximately  10\%. While this is a large error, it can be reduced by roughly one order of magnitude by the application of full {\sc supcode}. $\delta J$-{\sc supcode} carries larger error than the naive pulse as expected. Fig.~\ref{fig4}(b) shows the result for isotope-enriched Si. The gate errors are again much lower than those of natural Si: on a log-log plot the gate error of the na\"ive pulses reduces linearly from around 1\% for 3\% charge noise, to $10^{-5}$ for a charge noise of 0.1\%. In this range, the gate error is further reduced by an order of magnitude when {\sc supcode} pulses are used. Interestingly, the $\delta J$-{\sc supcode} reduces error more than the full {\sc supcode} which presumably arises from the simplicity of those pulses.

\begin{figure}[t]
    \centering
    \includegraphics[width=8cm, angle=0]{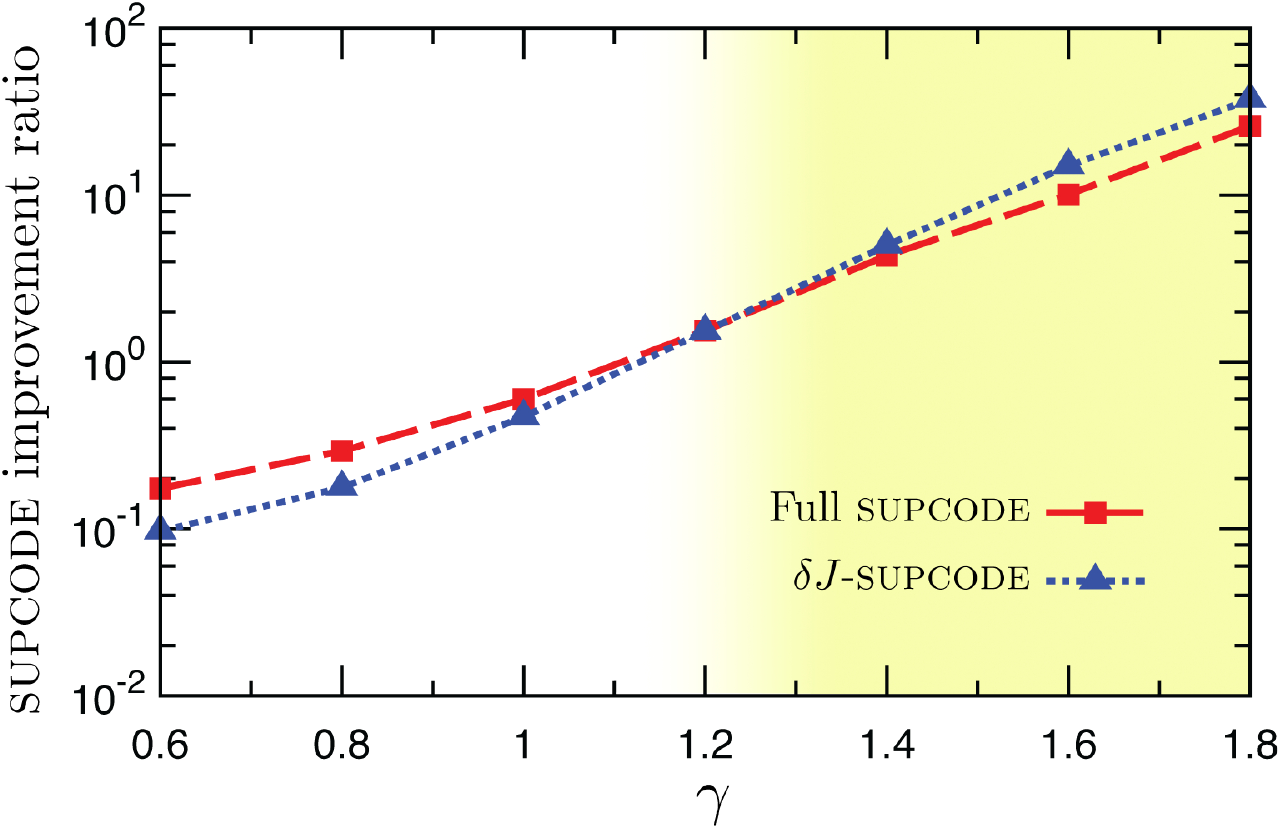}
    \caption{ The {\sc supcode} improvement ratio vs. charge noise exponent $\gamma$ when nuclear noise is absent. The result of full-{\sc supcode} is shown as the red dashed line and that of $\delta J$-{\sc supcode} is shown as the blue dotted line. The yellow area shows the regime where {\sc supcode} reduces error.}
    \label{fig5}
\end{figure}

To understand better the applicability of {\sc supcode} for different 1/f noises we show in Fig.~\ref{fig5} the results of the {\sc supcode} improvement ratio as a function of charge noise exponent, $\gamma$, when nuclear noise is absent. The improvement ratio is defined as the ratio between the average gate errors of na\"ive and {\sc supcode} pulses at asymptotically small charge noise. For a given exponent $\gamma$, this ratio is consistent for several different levels of small charge noise as indicated in the numerical calculation. Although the results shown in Fig.~\ref{fig4}(b) has a very small amount of nuclear noise, we can still see that the improvement ratio is approximately 10 and is consistent with the result of Fig.~\ref{fig5}. For large $\gamma$, the noise is concentrated at low frequency which is precisely what {\sc supcode} pulses are designed to correct, thus the improvement ratio increases exponentially as $\gamma$ increases. On the other hand, for smaller $\gamma$ the noises are more like white noise and performing {\sc supcode} would not only be useless for noise correction, but also introduces more error due to the longer gate time which allows the error to accumulate more. We therefore conclude that the {\sc supcode} pulses are useful for $\gamma>1.2$ and the improvement is much more pronounced if $\gamma$ is further increased.

In conclusion, we studied the performance of {\sc supcode} for both static noise and $1/f$ noise in Si (natural and isotope-enriched). We found that {\sc supcode} is useful for $1/f$ noise with exponent $\gamma>1.2$ in which case noise is concentrated at low frequency. In that regime, application of {\sc supcode} reduces the gate error in a substantial way.  We also demonstrated that $\delta J$-{\sc supcode} can perform comparably to the full {\sc supcode}, so it is particularly useful for isotope-enriched Si since it typically generates simpler and faster sequences. We note that the same protocol applies to two-qubit gates for single-spin qubits as well.\cite{Kalra.14} Our results clearly show that the extremely long coherence times of Si qubits pertain not only to information storage but also to information processing, if {\sc supcode} is applied. Thus, Si spin qubits may now be close to quantum error corrected fault tolerant quantum computing applications.

This work is supported by LPS-CMTC and IARPA. JPK acknowledges support from the UMBC Office of Research through an SRAIS award.

\appendix
\renewcommand{\theequation}{A-\arabic{equation}}
\setcounter{equation}{0}
\section{Parameters for $\delta J$-{\sc supcode}}\label{appham}

Here we provide parameters of the single-qubit Clifford gates for $\delta J$-{\sc supcode}, which is obtained from solving three coupled nonlinear equations (involving $\sigma_x$, $\sigma_y$, and $\sigma_z$ respectively) relevant to the charge noise only.  Note that $x$-rotations $R(\hat{x},-\pi/2)$, $R(\hat{x},\pi/2)$, and $R(\hat{x},\pi)$ do not need to turn on $J$ at all (therefore having no $\delta J$ error) and when charge noise is the only error source, the na\"ive pulses suffice.

The identity operation and the Hadamard gate can be made resistant to charge noise by using the following pulse sequence:
\begin{align}
&U\left(J,\pi+\frac{\phi}{2}\right)
U(j_3,\pi)
U(j_2,\pi)
U(j_1,\pi)
U(j_0,2\pi)\notag\\
&\times U(j_1,\pi)
U(j_2,\pi)
U(j_3,\pi)
U\left(J,\pi+\frac{\phi}{2}\right),
\label{eq:OnePieceSeq_djonly}
\end{align}
with parameters provided in Table~\ref{tab:numericonepiece_djonly}.

\begin{table}[h]
 \centering
 \begin{tabular}{|c||c|c|c|c|c|c|}
 \hline
 & $J$ & $\phi$ & $j_0$ & $j_1$ & $j_2$ & $j_3$ \\
 \hline
 \hline
 $I$ & 1 & 0 & 0.99998 & 9.9849 & 0.12167 & 10.000\\
 \hline
 $R(\hat{x}+\hat{z},\pi)$ & 1 & $\pi$ & 1.3604 & 10.000 & 0.18999 & 10.000\\
 \hline
 \end{tabular}
 \caption{Parameters of the $\delta J$-{\sc supcode} sequence, Eq.~\eqref{eq:OnePieceSeq_djonly}, appropriate for the identity operation and the Hadamard gate. Strengths of the exchange interactions are given in units of $h$.}\label{tab:numericonepiece_djonly}
\end{table}

Rotations $R(\hat{z},\pi/{2})$ and $R(\hat{z},\pi)$ can be done using the sequence
\begin{align}
&U(J=1,\pi)
U(j_4=0,\pi+\frac{\phi}{2})
U(j_3,\pi)
U(j_2,\pi)
U(j_1=0,\pi)\notag\\
&\times 
U(j_0,2\pi)U(j_1=0,\pi)
U(j_2,\pi)
U(j_3,\pi)
U(j_4=0,\pi+\frac{\phi}{2})\notag\\
&\times 
U(J=1,\pi),
\label{eq:zpulsecorr_djonly}
\end{align}
with parameters given in Table~\ref{tab:zpulsepara}.

\begin{table}[h]
 \centering
 \begin{tabular}{|c||c|c|c|c|c|c|c|}
 \hline
  & $\phi$ & $j_0$ & $j_1$ & $j_2$ & $j_3$ \\
 \hline
 \hline
 $R(\hat{z},\pi/{2})$ & $\pi/{2}$ &0.39727 & 9.9998 & 0.12151 & 9.9998\\
 \hline
 $R(\hat{z},\pi)$ & $\pi$ & 0.94156 & 10.000 & 0.12518 & 9.9994\\
 \hline
 \end{tabular}
 \caption{
 Parameters of the $\delta J$-{\sc supcode} sequence, Eq.~\eqref{eq:zpulsecorr_djonly}, appropriate for Clifford $z$-rotations. Strengths of the exchange interactions are given in units of $h$.
}\label{tab:zpulsepara}
\end{table}

$R(\hat{z},-\pi/{2})$ is done using the sequence
\begin{align}
&U(J=1,\pi)
U(j_5=0,\pi+\frac{\phi}{2})
U(j_4,\pi)
U(j_3,\pi)
U(j_2,\pi)\notag\\
&\times
U(j_1=0,\pi)
U(j_0,2\pi)
U(j_1=0,\pi)
U(j_2,\pi)
U(j_3,\pi)\notag\\
&\times
U(j_4,\pi)
U(j_5=0,\pi+\frac{\phi}{2})
U(J=1,\pi)
\label{eq:zpulsecorr_djonly_2}
\end{align}
with parameters given in Table~\ref{tab:zpulsepara_2}.

\begin{table}[h]
 \centering
 \begin{tabular}{|c||c|c|c|c|c|c|c|}
 \hline
  & $\phi$ & $j_0$ & $j_1$ & $j_2$ & $j_3$ & $j_4$\\
 \hline
 \hline
 $R(\hat{z},-\pi/{2})$ & $-\pi/{2}$ & 9.9090 & 0.26697 & 9.1544 &
0.68434 & 10.000 \\
 \hline
 \end{tabular}
 \caption{
 Parameters of the $\delta J$-{\sc supcode} sequence, Eq.~\eqref{eq:zpulsecorr_djonly_2}, appropriate for Clifford $ R(\hat{z},-\pi/2)$ rotation. Strengths of the exchange interactions are given in units of $h$.
}\label{tab:zpulsepara_2}
\end{table}

A general rotation around any arbitrary axis can be achieved by the following sequence
\begin{align}
&U(J=0,\phi_a)
U(J=1,\pi)
U(j_4=1,\pi+\theta_4)
U(j_3,\pi)\notag\\
&\times
U(j_2,\pi)
U(j_1,\pi)
U(j_0,2\pi)
U(j_1,\pi)
U(j_2,\pi)
U(j_3,\pi)\notag\\
&\times
U(j_4=1,\pi-\theta_4)
U(J=0,\phi_b)
U(J=1,\pi)
U(J=0,\phi_c).
\label{eq:FivePieceSeq_djonly}
\end{align}
with parameters given in Table~\ref{tab:genpulsepara}.

\begin{table*}
 \centering
 \begin{tabular}{|c||c|c|c|c|c|c|c|c|c|c|c|}
 \hline
  & $j_0$ & $j_1$ & $j_2$ & $j_3$ & $\theta_4$ & $\phi_a$ &  $\phi_b$ & $\phi_c$ \\
 \hline
 \hline
 $R(\hat{y},-\pi/{2})$ & 10.000 & 0.41333 & 10.000 & 0.24570 &  $\arctan[(4+\sqrt{2}\pi)/(-2\sqrt{2}+\pi)] $ & $3\pi/2$ & $3\pi/2$ & $\pi/2$ \\
 \hline
 $R(\hat{y},\pi/{2})$ &10.000 & 0.41333 & 10.000 & 0.24570 &  $\arctan[(4+\sqrt{2}\pi)/(-2\sqrt{2}+\pi)] $ & $5\pi/2$ & $3\pi/2$ & $3\pi/2$\\
 \hline
 $R(\hat{y},\pi)$ & 1.2144 & 10.000 & 0.26486 & 10.000 &  $-2\arctan[(\pi+\sqrt{16+\pi^2})/4]$ & $3\pi/2$ & $\pi$ & $\pi/2$\\
 \hline
 \hline
$R(\hat{x}-\hat{z},\pi)$ & 10.000 & 0.41333 & 10.000 & 0.24570 &  $\arctan[(4+\sqrt{2}\pi)/(-2\sqrt{2}+\pi)] $ &  $\pi/2$ & $3\pi/2$ & $\pi/2$\\
\hline
$R(\hat{x}+\hat{y},\pi)$ & 9.9957 & 0.30460 & 9.9964 & 0.27306 & $\arctan[(4-\sqrt{2}\pi)/(2\sqrt{2}+\pi)]$ & 0 & $\pi/2$ & $3\pi$\\
\hline
$R(\hat{x}-\hat{y},\pi)$ & 9.9957 & 0.30460 & 9.9964 & 0.27306 &  $\arctan[(4-\sqrt{2}\pi)/(2\sqrt{2}+\pi)]$  & $\pi$ & $5\pi/2$ & $2\pi$\\
\hline
$R(\hat{y}+\hat{z},\pi)$ & 10.000 & 0.33692 & 9.9951 & 0.34373 & $\arctan(4/\pi)$ & $7\pi/2$ & $\pi$ & $2\pi$\\
\hline
$R(\hat{y}-\hat{z},\pi)$ & 10.000 & 0.33692 & 9.9951 & 0.34373 & $\arctan(4/\pi)$ & $\pi/2$ & $\pi$ & 0\\
\hline
\hline
$R(\hat{x}+\hat{y}+\hat{z},2\pi/3)$ & 8.5686 & 0.29793 & 10.000 & 0.27073 & $\arctan[(4-\sqrt{2}\pi)/(2\sqrt{2}+\pi)]$ & 0 & $\pi/2$ & $\pi/2$ \\
\hline
$R(\hat{x}+\hat{y}+\hat{z},4\pi/3)$ & 10.000 & 0.41333 & 10.000 & 0.24570 & $\arctan[\sqrt{2}+8(2\sqrt{2}+\pi)/(\pi^2 -8)]$ & $7\pi/2$ & $7\pi/2$ & $2\pi$\\
\hline
$R(\hat{x}+\hat{y}-\hat{z},2\pi/3)$ &  10.000 & 0.41333 & 10.000 & 0.24570 &$\arctan[(4+\sqrt{2}\pi)/(-2\sqrt{2}+\pi)]$ & $5\pi/2$ & $3\pi/2$ & $4\pi$\\
\hline
$R(\hat{x}+\hat{y}-\hat{z},4\pi/3)$ & 9.9957 & 0.30460 & 9.9964 & 0.27306 &  $\arctan\{[-8\sqrt{2}+\pi(8-\sqrt{2}\pi)]/(\pi^2 -8)\}$ & 0 & $\pi/2$ & $3\pi/2$\\
\hline
$R(\hat{x}-\hat{y}+\hat{z},2\pi/3)$ & 8.7287 & 0.29878 & 10.000 & 0.27104 & $\arctan\{[-8\sqrt{2}+\pi(8-\sqrt{2}\pi)]/(\pi^2 -8)\}$ & $\pi/2$ & $5\pi/2$ & $2\pi$\\
\hline
$R(\hat{x}-\hat{y}+\hat{z},4\pi/3)$ & 10.000 & 0.41333 & 10.000 & 0.24570 & $\arctan[(4+\sqrt{2}\pi)/(-2\sqrt{2}+\pi)]$ & $2\pi$ & $7\pi/2$ & $3\pi/2$ \\
\hline
$R(-\hat{x}+\hat{y}+\hat{z},2\pi/3)$ & 9.9957 & 0.30460 & 9.9964 & 0.27306 & $\arctan[(4-\sqrt{2}\pi)/(2\sqrt{2}+\pi)]$ & $3\pi/2$ & $\pi/2$ & $2\pi$ \\
\hline
$R(-\hat{x}+\hat{y}+\hat{z},4\pi/3)$ & 10.000 & 0.41333 & 10.000 & 0.24570 & $\arctan[\sqrt{2}+8(2\sqrt{2}+\pi)/(\pi^2 -8)]$ & $4\pi$ & $7\pi/2$ & $5\pi/2$\\
\hline
 \end{tabular}
 \caption{
Parameters of the $\delta J$-{\sc supcode} sequence, Eq.~\eqref{eq:FivePieceSeq_djonly}, appropriate for all remaining Clifford gates. Strengths of the exchange interactions are given in units of $h$.}\label{tab:genpulsepara}
\end{table*}

In Table~\ref{tab:length} we compare the pulse length of the na\"ive pulse ($L_\mathrm{naiv}$), full {\sc supcode} ($L_\mathrm{full}$) and $\delta J$-{\sc supcode} ($L_{\delta J}$) in unit of $\left(h^{-1}\right)$. From the  ratio $L_{\delta J}/L_\mathrm{full}$ we clearly see that $\delta J$-{\sc supcode} is approximately $30\%\sim50\%$ shorter than the full {\sc supcode} (except the $x$-rotations), demonstrating its advantage to be implemented in isotope-enriched Si.

\begin{table*}[h]
 \centering
 \begin{tabular}{|c||c|c|c|c||c||c|c|c|c|}
 \hline
 & $L_\mathrm{naiv}$ & $L_\mathrm{full}$ & $L_{\delta J}$ & $L_{\delta J}/L_\mathrm{full}$ & & $L_\mathrm{naiv}$ & $L_\mathrm{full}$ & $L_{\delta J}$ & $L_{\delta J}/L_\mathrm{full}$  \\
 \hline
 \hline
$I$ &   4.443& 30.92& 16.37& 53.0\% & $R(\hat{x}+\hat{y},\pi)$ & 9.155 & 57.34 & 33.20 & 57.9\%\\
\hline
$R(\hat{x},-\pi/{2})$ &   4.712& 29.84& 4.712& 15.8\%& $R(\hat{x}-\hat{y},\pi)$ & 9.155 & 63.72 & 39.49 & 62.0\%\\
\hline
$R(\hat{x},\pi/{2})$ &  1.571& 42.07& 1.571&  3.73\% & $R(\hat{y}+\hat{z},\pi)$ & 12.30 & 66.33 & 42.45 & 64.0\%\\
\hline
$R(\hat{x},\pi)$ & 3.142& 28.10& 3.142& 11.2\% & $R(\hat{y}-\hat{z},\pi)$ & 9.155 & 53.15 & 26.74 & 50.3\%\\
\hline
$R(\hat{y},-\pi/{2})$ & 15.44& 56.34& 33.04&  58.6\%&  $R(\hat{x}+\hat{y}+\hat{z},2\pi/3)$ & 7.584 & 50.68 & 25.47 & 50.2\% \\
\hline
$R(\hat{y},\pi/{2})$ & 12.30&  60.89& 39.32& 64.6\%& $R(\hat{x}+\hat{y}+\hat{z},4\pi/3)$ & 13.87 & 71.13 & 50.32 & 70.7\%\\
\hline
$R(\hat{y},\pi)$ & 13.87& 56.62&  29.63& 52.3\%& $R(\hat{x}+\hat{y}-\hat{z},2\pi/3)$ &  10.73 & 72.95 & 47.18 & 64.7\%\\
\hline
$R(\hat{z},-\pi/{2})$ &   9.155& 38.77& 22.35& 57.6\%& $R(\hat{x}+\hat{y}-\hat{z},4\pi/3)$ & 10.73 & 53.40 & 28.49 & 53.4\%\\
\hline
$R(\hat{z},\pi/{2})$ &  6.014&  47.66& 25.62& 53.8\%& $R(\hat{x}-\hat{y}+\hat{z},2\pi/3)$ & 7.584 & 62.24 & 38.02 & 61.1\%\\
\hline
$R(\hat{z},\pi)$ &   7.584& 50.96& 25.93& 50.9\%& $R(\hat{x}-\hat{y}+\hat{z},4\pi/3)$ & 13.87 & 65.48 & 44.04 & 67.3\%\\
\hline
$R(\hat{x}+\hat{z},\pi)$ &   2.221& 28.76&  17.81& 61.9\%&  $R(-\hat{x}+\hat{y}+\hat{z},2\pi/3)$ & 10.73 & 59.16 & 34.77 & 58.8\%\\
\hline
$R(\hat{x}-\hat{z},\pi)$ &  12.30& 53.50& 29.90&  55.9\%& $R(-\hat{x}+\hat{y}+\hat{z},4\pi/3)$ & 10.73 & 71.91 & 53.46 & 74.3\%\\
\hline
 \end{tabular}
 \caption{
Pulse length of the na\"ive pulse ($L_\mathrm{naiv}$), full {\sc supcode} ($L_\mathrm{full}$) and $\delta$ {\sc supcode} ($L_{\delta J}$) in unit of $\left(h^{-1}\right)$. The ratio $L_{\delta J}/L_\mathrm{full}$, showing the reduction in length of $\delta J$-{\sc supcode} from the full one, is also presented.}\label{tab:length}
\end{table*}

\end{document}